# A Directly-Written Monolithic Waveguide-Laser Incorporating a DFB Waveguide-Bragg Grating

**Graham D. Marshall, Peter Dekker, Martin Ams, James A. Piper, and Michael J. Withford**
*Centre for Ultrahigh bandwidth Devices for Optical Systems, MQ Photonics Research Centre,*
*Department of Physics, Macquarie University, NSW 2109, Australia*
*graham@ics.mq.edu.au*

We report the fabrication and performance of the first C-band directly-written monolithic waveguide-laser. The waveguide-laser device was created in an Erbium and Ytterbium doped phosphate glass host and consisted of an optical waveguide that included a distributed feedback Bragg grating structure. The femtosecond laser direct-write technique was used to create both the waveguide and the waveguide-Bragg grating simultaneously and in a single processing step. The waveguide-laser was optically pumped at approximately 980 nm and lased at 1537nm with a bandwidth of less than 4 pm.

The use of ultrafast lasers to inscribe photonic components in optical media, the so called direct-write technique, is a research field that is attracting considerable attention. The technique enables optical waveguide devices to be written in active and passive glasses through the use of a tightly focused femtosecond laser beam that is scanned through an optical medium of interest. By translating the optical medium on computer controlled translation stages arbitrary waveguide designs can be rapidly fabricated. The non-linear materials-interaction processes that are strongest at the focus of the writing-laser can yield a region of local positive refractive index change that forms an optical waveguide [1]. This process of non-destructive material modification is most commonly applied to amorphous glass hosts in which positive refractive index changes can often be achieved. A wide range of devices have been created using the direct-write technique, this includes splitters [2], interferometers and gratings [3, 4]. The technique has also been used to create amplifying waveguide regions that have formed the basis for waveguide-lasers however these devices have always relied upon fiber-Bragg gratings to act as extra-waveguide reflectors [5, 6]. Nevertheless the performance of these devices is impressive with output powers as high as 80 mW reported with slope efficiencies of 21% [5]; results that exemplify the potential of the direct-write technique to create devices for integrated optical systems. In this Letter we present the first report on the creation of a directly-written *monolithic* waveguide-laser device which unlike previous waveguide-laser designs incorporates an intra-waveguide laser cavity mirror. This mirror was in the form of a distributed feedback (DFB) waveguide-Bragg grating (WBG) that was coherent over the length of the device. The resulting waveguide-laser (WGL) was double-end pumped through optical fiber wavelength-division multiplexers (WDMs) that split the WGL light field from the pump light. Our simple laser fabrication technique differs from previously reported and more complicated methods that required lithographic, ion-exchange, interferometric grating writing and reactive etching processes [7], combined interferometrically written waveguide-Bragg gratings and plane mirror architectures [8], or used a holographic system to create a string of non phased-aligned microgratings [9]. Distributed feedback lasers are characterized by their narrow linewidth, single frequency output [10] and our use of this architecture well matches the capabilities of the direct-write technique with the requirements of modern integrated optical systems.

The laser used to write the waveguide-laser device was a 1 kHz repetition rate, 120 fs pulse length, 800 nm regeneratively amplified Ti:Sapphire laser which was focused into the glass sample using a 20×, 0.35 effective NA microscope objective. The writing-laser beam was circularly polarized because this polarization has been demonstrated to create waveguides with the lowest propagation losses in other host materials [11]. A slit was placed in front of the microscope objective to modify the shape of the laser focus and create circularly symmetric waveguides as reported in [12]. The glass sample was a 2% (by wt.) Erbium and 4% Ytterbium co-doped 'QX' phosphate glass host (Kigre, USA) that was mounted atop an air bearing based three-axis translation stage system (Aerotech, USA). The 20 mm long glass sample was translated at 25 μm/s through the focused writing-laser beam. In order to create the WBG structure a similar technique to that reported in [4] was used. The writing-laser was 100% intensity modulated with a 50:50 mark-space ratio whilst the glass sample was being translated thereby creating a waveguide formed by segments of exposed glass with an approximately 500 nm pitch. This corresponded to a $1^{st}$ order Bragg grating structure in the glass. Modulation of the writing-laser intensity was conveniently achieved by interrupting the TTL level trigger signal to the regen' amplifier's Pockels cells. The pulse energy used to create the waveguide-Bragg grating based laser device was 1.6 μJ as measured after the slit. The glass sample was ground and polished back by 150 μm (at each end facet) after device fabrication to remove the distorted regions of waveguide where the writing-laser beam entered and exited the sample. Fig. 1 shows a transmission differential interference contrast (DIC) micrograph of the fabricated first order waveguide-Bragg grating structure. The modified waveguiding region was approximately 7 μm in diameter and clearly shows a 500 nm periodic refractive index variation. The waveguide supported a single transverse mode in the C-band indicating that, if assumed to be of step-index profile, the maximum refractive index difference between the waveguide and the bulk glass was $2.3 \times 10^{-3}$. The highly non-linear interaction between the writing-laser focus and the glass material allowed the creation of the sub-wavelength grating features observed in Fig. 1. However due to the physical limit of diffraction imposed on the writing-laser beam's focus, the grating periods cannot be assumed to have a square-wave refractive index profile that replicates the writing-laser intensity modulation. It is more likely that the refractive index profile is approximated by a triangular-cum-sinusoidal form similar to that observed in the inset of Fig. 1. wherein a graph of the transmission-DIC micrograph intensity is shown.

The pumping and diagnostic arrangement shown in Fig. 2 was used to measure the performance of the WGL which was double-end pumped using two pump laser diodes with 980 nm and 976 nm output wavelengths. The 976 nm laser diode was connected to a WDM via an optical isolator. The output fiber from the WDMs (Corning Hi-980) supported too small a mode-field diameter at the pump wavelength (4.2 μm) to couple efficiently to the WGL. To overcome the loss associated with this mismatch between the waveguide and fiber mode-field diameters approximately 250 μm long sections of graded index optical fiber (GIF625) were fusion spliced to the Hi-980 pump fiber tips and acted as mode-field converters in a manner similar to that reported in [13]. To facilitate stable operation of the DFB WGL an estimated π/2 phase shift was induced at the center of the WBG using a small external heater (a surface mount resistor placed in contact with the host glass). The application of a heat to create a centre-grating phase shift in a fiber DFB laser device has been reported previously and enables single-longitudinal mode operation at a wavelength at the centre of the Bragg grating stop-band [10].

The waveguide-Bragg grating reflection and transmission characteristics were studied using a swept laser source and a three port circulator; these were connected to the device through the WDMs. Because the waveguide-Bragg grating resonance under test coincided with the quasi three-level laser Erbium transition any C-band optical measurement performed on the Bragg grating was inherently subject to absorption of the probe light. Therefore measurements of the reflection and transmission efficiencies can not be made exactly and are lower and upper bounds on these quantities respectively. The reflection and transmission characteristics of the waveguide-Bragg grating without the thermally induced central phase shift are shown in Fig. 3; the graphs are not corrected for WDM, coupling or the unknown waveguide propagation and absorption losses. The waveguide-Bragg grating has one sharp Bragg resonance with a FWHM of 140 pm indicating the excellent coherence or phase stability of the grating. The transmission spectrum of the grating is characterized by material absorption losses on the long wavelength side and losses to the continuous cladding of the WBG on the short wavelength side of the Bragg resonance. In reflection the grating is characterized by weak scattering reflections at wavelengths outside of the strongly reflecting resonance peak.

The monolithic waveguide-laser device was pumped with up to 710 mW of light from 976 nm and 980 nm pump diodes. The maximum power from the 976 nm and 980 nm pump diodes at the WGL were 364 mW and 346 mW respectively. The threshold pump power for lasing was 639 mW and the maximum recorded output power from the WGL device was 0.37 mW (the summed output power emitted symmetrically from each output facet). The output power of the WGL was limited by and increased monotonically with the applied pump power. The output from the WGL observed with a 10 pm slit-width optical spectrum analyzer is shown in Fig. 4. The WGL output spectrum displays a single peak in intensity with a width that is limited by the instrument's resolution and a greater than 50 dB side-mode suppression ratio (SMSR). The laser wavelength measured by a scanning Michelson wavemeter was 1537.627 nm and the same instrument provided an estimate of the maximum laser linewidth (based on coherence length) of 4 pm. The centre laser wavelength was stable to $\pm 3$ pm over a measurement period of 5 minutes and the source of this drift was expected to be due to variations in the temperature of the waveguide structure induced by changing laboratory environmental conditions. The narrow line-width output, good wavelength stability and high SMSR makes this WGL a suitable source for dense wavelength division multiplexing applications. The wavelength of this design of laser is specified by the period of the WBG (subject to the gain-bandwidth constraints of the host material) and the refractive index of the medium. Measurements of the reflection spectra from WBGs written with known periods indicate that the refractive index of the host material used in this study was 1.5398 at 1535 nm and 1.5391 at 1560 nm. Given these data WGLs created using our technique could be easily aligned with the 100 GHz to 25 GHz International Telecommunications Union (ITU) frequency grids. Furthermore it is expected that the WGL device wavelength could be thermally tuned wherein the host material's coefficient of optical path length with temperature would determine $d\lambda/dT$.

This Letter is the first report of a directly-written monolithic waveguide-laser and of the meaningful integration of two direct-write platforms, namely a WBG and an amplifier, to form a genuinely integrated photonic device. Our technique for creating these devices is extremely flexible and enables the creation of narrow linewidth lasers in bulk optical materials without the need for external components. A simple computer code controls the laser, translation stage system and frequency generator used for creating the WBG. The ratio of the sample translation speed to the writing-laser modulation frequency controls the WBG period and we proposed that it would be possible to modify the refractive index contrast of the grating periods and therefore the grating reflectivity by altering the mark-space ratio of the modulation square wave applied to the writing-laser (albeit with a small change in the WBG effective refractive index). This technique for WGL fabrication is applicable to all doped glass hosts that can be femtosecond-laser processed to create a waveguiding region and is limited only by the period of the shortest wavelength grating that can be realized in this fashion. The resulting devices are naturally compatible with existing waveguide devices such as splitters, amplifiers and interferometers. The relatively high threshold power (as compared to WGLs based on extra-waveguide cavity mirrors) and low optical efficiency of this WGL indicates that there is potential to improve the performance of the device by refining the WBG grating characteristics, improving the mode coupling between the waveguide-laser and launch fibers and optimizing the active device length and dopant concentrations. Further device investigations will include measurements of the laser relative intensity noise, Fabry-Pérot interferometer analysis of the laser spectrum and further waveguide-Bragg grating characterization.

This work was supported by the Australian Research Council through their Centres of Excellence and LIEF Programs.

Figures

Fig. 1. Transmission DIC micrograph of the 1$^{st}$ order WBG. The inset shows a section of the WBG at 3× magnification. Overlaid on the inset is a graph of the grating image intensity.

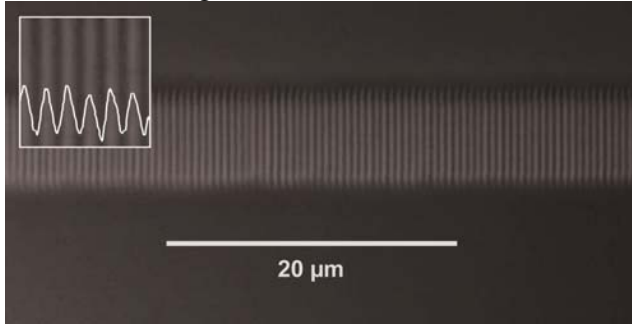

Fig. 2. Waveguide laser pumping and diagnostic setup.

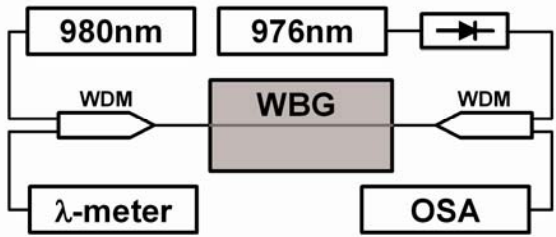

Fig. 3. Reflection and transmission spectra of the un-pumped waveguide-Bragg grating.

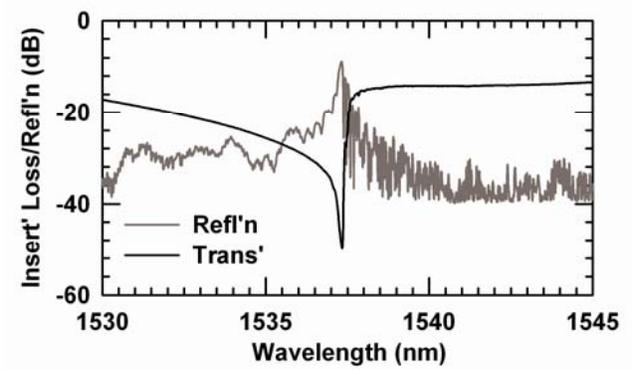

Fig. 4. Output spectrum of the waveguide laser at maximum pump power. The ordinate axis scale is referenced to the laser peak.

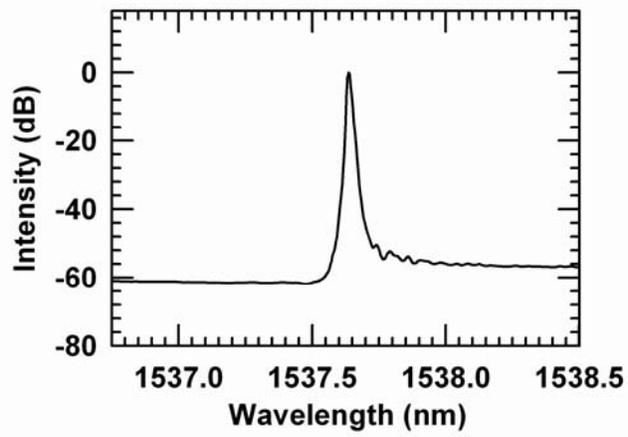